\documentclass[sigconf]{acmart}

\AtBeginDocument{%
  \providecommand\BibTeX{{%
    \normalfont B\kern-0.5em{\scshape i\kern-0.25em b}\kern-0.8em\TeX}}}

\renewcommand\footnotetextcopyrightpermission[1]{}

\acmConference[HCI-TERRA: HCI Towards EnviRonmentally Responsible AI, CHI '26]{ACM Conference}{April 16, 2026}{Barcelona, Spain}%

\begin{document}

\title{An HCI Perspective on Sustainable GenAI Integration in Architectural Design Education}
\renewcommand{\shorttitle}{Sustainable GenAI Integration in Architectural Design Education}

\author{Alex Binh Vinh Duc Nguyen}
\email{alexbvd.nguyen@uantwerpen.be}
\orcid{0000-0001-5026-474X}
\affiliation{
  \institution{University of Antwerp}
  \city{Antwerp}
  \country{Belgium}
}
\affiliation{
  \institution{KU Leuven}
  \city{Leuven}
  \country{Belgium}
}


\begin{abstract}
Generative AI (genAI) is increasingly influencing architectural design practice and is expected to affect, or even transform, the profession, even though its benefits and costs remain unresolved. In response, design schools are increasingly integrating genAI into their curricula. Yet this integration creates a paradox: critical engagement with genAI often requires increased use of the tools in question, despite limited methods for estimating their environmental cost in teaching contexts. 
In this paper, we argue that HCI offers a useful methodological lens for addressing this tension. We propose three HCI-informed directions for more sustainable genAI integration in architectural education: contextual eco-feedback, participatory stakeholder scoping, and reframing data centres as an interdisciplinary focus. We therefore argue that genAI should be understood not only as a new architectural design tool, but also as a socio-technical process that architectural education, and design education in general, must engage with critically.
\end{abstract}



\maketitle

\section{GenAI in architectural design: State-of-the-art}

\subsection{Practice}

Generative AI (genAI) technologies are providing architects with new tools aimed at supporting design tasks. Recent research suggests that these tools can offer situationally relevant inspiration during early conceptual phases~\cite{vissers-similon_classification_2025,paananen_using_2024}, automate non-creative tasks~\cite{han_automatic_2024,jang_automated_2024}, and, in some cases, learn aspects of an architect's design style to propose plausible alternatives~\cite{Lee2024}. As genAI capabilities continue to develop rapidly, their influence on architectural practice is increasingly difficult to ignore.

At the same time, genAI is also framed as a disruptive force that may challenge even established roles in the design-build process~\cite{lyu_generative_2024}. For example, one recent economic analysis estimates that up to 37\% of tasks in the architecture and engineering industry could be exposed to automation by genAI~\cite{hatzius2023potentially}. However, many architects argue that these tools still cannot replace the critical judgement, empathy, cultural sensitivity, and situated intuition required in architectural design~\cite{moral-andres_can_2024}. Studies of AI-generated architectural outputs also identify ethical and professional concerns, including socio-cultural bias in training data~\cite{moral-andres_generative_2024}, limited reliability and explainability of design decisions~\cite{sonkor_using_2025}, and unresolved questions of liability for safety and quality control~\cite{liang_ethics_2024}. In addition, the environmental impact of genAI, although widely discussed in public and academic discourse, remains difficult to measure in practice~\cite{ITU2025MeasuringWhatMatters,OECD2022AIFootprint,IEA2025EnergyAI}. Taken together, these developments indicate that while genAI is likely to affect architectural design substantially, the balance between its benefits and costs remains unresolved.

\subsection{Education}
In response to these developments, architectural and design schools have started to introduce genAI-related content into their curricula through workshops, elective courses, and studio exercises~\cite{Onatayo2024AECReview}. These initiatives aim to help students engage with both the opportunities and the limitations of genAI in design practice~\cite{Jin2024AIAssistedCourse,Ozorhon2025AIaADS}. Given the socio-technical complexity of genAI, such integration will likely become more common as educators seek to prepare future architects to work critically with genAI tools, or to consciously resist them while still understanding their ethical, social, and environmental implications~\cite{Adebayo2025BuiltEnvAcademics}.

However, this integration introduces a practical paradox. Teaching students to critically evaluate genAI often also requires them to use genAI tools more intensively, which inevitably contributes to the environmental impact of these tools. At present, there is still no widely adopted method that guides architectural educators in estimating the environmental cost of a genAI-focused course or workshop~\cite{ITU2025MeasuringWhatMatters,OECD2022AIFootprint}. As a result, educators are often required to make curriculum decisions based on personal beliefs and attitudes toward the technology, which have been shown to vary widely among architects and design professionals~\cite{Adebayo2025BuiltEnvAcademics, fernberg_problematizing_2024}.

\section{HCI as a Methodological Lens for Sustainable genAI Architectural Education}

In this context, HCI research offers a useful methodological lens for informing genAI architectural education, and more broadly, for shaping the future architectural profession. This is because HCI provides established approaches for analysing socio-technical systems, designing interactions that support reflective decision-making, and engaging affected communities through participatory methods. For example, we propose three directions below to illustrate how HCI knowledge can contribute to genAI architectural education.

\subsection{Providing proactive and contextual eco-feedback}
At present, one of the most common uses of genAI in architectural design is image generation~\cite{Onatayo2024AECReview}. Because architectural images must satisfy conceptual, structural, and contextual expectations, designers often go through many iterations of prompting and reconfiguration before reaching a satisfactory outcome~\cite{paananen_using_2024}. This iterative workflow can produce substantial computational overhead, but this cost is rarely visible during the design process.

HCI research can address this gap by developing proactive eco-feedback mechanisms~\cite{FroehlichFindlaterLanday2010EcoFeedback}. Such mechanisms should not only visualise the estimated environmental cost of a generation task before execution, but may also support designers in reducing unnecessary iterations, for example, by helping them refine prompts more effectively at earlier stages. Importantly, this feedback should be contextualised rather than presented as an isolated metric. For example, the environmental cost of a genAI image workflow should be compared against typical alternatives in architectural practice, such as manual 3D modelling and rendering workflows, while accounting for likely iteration counts, output quality requirements, and the stage of design development. Without this context, eco-feedback risks becoming informative but not actionable.

\subsection{Engaging stakeholders to scope educational responsibility}
The integration of genAI into architectural education is often justified by the assumption that genAI will substantially affect the profession. However, it remains unclear how and to what extent this impact is unfolding. This uncertainty weakens the grounding of genAI education and can lead to curriculum design that depends heavily on individual educators' attitudes or on short-term technological trends.

HCI participatory methods can help address this problem by engaging stakeholders in defining what responsible and sustainable genAI education should include. These efforts should involve not only stakeholders within the architectural sector, such as architects, engineers, contractors, and clients, but also human-centred AI researchers, AI developers, as well as students and educators. Participatory approaches can help clarify which values are at stake, who is affected by specific educational choices, and what forms of competence should be prioritised in the curriculum. An important open question remains the scope of responsibility: it is still not clear which stakeholders and communities are directly or indirectly affected by genAI integration in architectural design. This scoping question should be treated as a research problem in its own right rather than assumed in advance.

\subsection{Reframing data centre design as an interdisciplinary focus}
Discussion of genAI in architecture often focuses on how AI tools may change design practice, but far less attention is given to how architectural practice and education might contribute to a more sustainable future of AI infrastructure itself. Data centres are central to genAI systems~\cite{IEA2025EnergyAI}, yet they remain marginal in many architecture curricula. This limits architects' ability to contribute meaningfully to the design of these facilities as environmental, urban, and social infrastructures.

Although there are emerging initiatives in more responsible data centre design, for example, recovering waste heat, managing water use more carefully, or integrating data centres into urban public spaces~\cite{Arup2024SustainableDataCentre,Arup2024WaterDataCentre}; architects still often lack the interdisciplinary knowledge required to work on these projects. We thus propose that this is where HCI and architecture can productively intersect. HCI can contribute methods for stakeholder engagement, operational transparency, and communication of infrastructural impacts to affected communities, while architectural education can contribute spatial, material, and urban design expertise. As digital and physical infrastructures become increasingly entangled, this interdisciplinary framing is likely to become more important.

\section{Conclusion}
GenAI development is already reshaping architectural practice and is beginning to reshape architectural education. However, the direction and consequences of these processes remain uncertain. While current educational responses to genAI in architecture are necessary, they also contribute to its environmental impact, as critical engagement with genAI often depends on increased use of the very systems in question. 
In this context, HCI research offers an applicable set of methods and design approaches for responding to this tension, particularly through contextual eco-feedback, participatory scoping of educational responsibility, and interdisciplinary work on AI infrastructure. Rather than treating genAI as a new design tool to adopt or reject, architectural education may benefit from treating it as a socio-technical process that must be analysed, designed, and governed critically.


\bibliographystyle{ACM-Reference-Format}
\balance
\bibliography{sample-base}


\end{document}